\documentclass[12pt]{article}
\usepackage{latexsym,amssymb}

\newcommand{\eq}[1]{(\ref{#1})}

\newcommand{\be}{\begin{equation}}
\newcommand{\ee}{\end{equation}}
\newcommand{\bea}{\begin{eqnarray}}
\newcommand{\eea}{\end{eqnarray}}

\newcommand{\ba}{\begin{eqnarray}}
\newcommand{\ea}{\end{eqnarray}}

\def\ibar{{\bar \imath}}
\def\jbar{{\bar \jmath}}

\def\Im{{\rm Im }}
\def\Re{{\rm Re }}
\def\IP{\relax{\rm I\kern-.18em P}}

\font\cmss=cmss10 \font\cmsss=cmss10 at 7pt
\def\twomat#1#2#3#4{\left(\matrix{#1 & #2 \cr #3 & #4}\right)}
\def\inbar{\vrule height1.5ex width.4pt depth0pt}
\def\IC{\relax\,\hbox{$\inbar\kern-.3em{\rm C}$}}
\def\IG{\relax\,\hbox{$\inbar\kern-.3em{\rm G}$}}
\def\IB{\relax{\rm I\kern-.18em B}}
\def\ID{\relax{\rm I\kern-.18em D}}
\def\IL{\relax{\rm I\kern-.18em L}}
\def\IF{\relax{\rm I\kern-.18em F}}
\def\IH{\relax{\rm I\kern-.18em H}}
\def\II{\relax{\rm I\kern-.17em I}}
\def\IN{\relax{\rm I\kern-.18em N}}
\def\IP{\relax{\rm I\kern-.18em P}}
\def\IQ{\relax\,\hbox{$\inbar\kern-.3em{\rm Q}$}}
\def\bfzero{\relax\,\hbox{$\inbar\kern-.3em{\rm 0}$}}
\def\IK{\relax{\rm I\kern-.18em K}}
\def\IG{\relax\,\hbox{$\inbar\kern-.3em{\rm G}$}}
 \font\cmss=cmss10 \font\cmsss=cmss10 at 7pt
\def\IR{\relax{\rm I\kern-.18em R}}
\def\ZZ{\relax\ifmmode\mathchoice
{\hbox{\cmss Z\kern-.4em Z}}{\hbox{\cmss Z\kern-.4em Z}}
{\lower.9pt\hbox{\cmsss Z\kern-.4em Z}} {\lower1.2pt\hbox{\cmsss
Z\kern-.4em Z}}\else{\cmss Z\kern-.4em Z}\fi}
\def\bfone{\relax{\rm 1\kern-.35em 1}}

\def\ii{{\mathrm i}}
\def\e{{\mathrm e}}

\def\IU{\relax\,\hbox{$\inbar\kern-.3em{\rm U}$}}

\def\be{\beta}

\def\part{\partial}

\def\square{{\,\lower0.9pt\vbox{\hrule \hbox{\vrule height 0.2 cm
\hskip 0.2 cm \vrule height 0.2 cm}\hrule}\,}}

\def\bfone{\relax{\rm 1\kern-.35em 1}}
\font\cmss=cmss10 \font\cmsss=cmss10 at 7pt

\def\cL{{\cal L}} \def\cM{{\cal M}}
\def\cN{{\cal N}}

\def\bar{\overline}

\def\Coe#1.#2.{\frac{#1}{ #2}}

\def\coe#1.#2.{\relax{\textstyle {#1 \over #2}}\displaystyle}

\def\to{\rightarrow}
\def\notin{\hbox{{$\in$}\kern-.51em\hbox{/}}}

\def\IE{\relax{{\rm I\kern-.18em E}}}

\def\IGam{\relax{{\rm I}\kern-.18em \Gamma}}

\def\inbar{\vrule height1.5ex width.4pt depth0pt}
\def\bfzero{\relax{\rm I\kern-.18em 0}}
\def\bfone{\relax{\rm 1\kern-.35em 1}}
\def\twomat#1#2#3#4{\left(\begin{array}{cc}
\end{array}
\right)}

\def\cL{{\cal L}} \def\cM{{\cal M}}
\def\cN{{\cal N}}

\begin{document}

  \begin{center}{\LARGE \bf  First Order Description of Black Holes in Moduli Space}
\vskip 1.5cm
{L. Andrianopoli$^{1,2}$,  R. D'Auria$^2$, E. Orazi$^{2}$ and M. Trigiante$^2
$}   \end{center}
 \vskip 3mm
\noindent
{\small
$^1$ Centro E. Fermi, Compendio Viminale,
I-00184 Rome, Italy and CERN PH-TH Division, CH 1211 Geneva 23, Switzerland
  \\
$^2$
Dipartimento di Fisica,
  Politecnico di Torino, Corso Duca degli Abruzzi 24, I-10129
  Turin, Italy and Istituto Nazionale di Fisica Nucleare (INFN)
  Sezione di Torino, Italy
 }
 \footnote{

 \texttt{laura.andrianopoli@polito.it};

 \texttt{riccardo.dauria@polito.it};

  \texttt{emanuele.orazi@polito.it};

   \texttt{mario.trigiante@polito.it}.}

\vfill
\vskip 1,5 cm

\begin{abstract}
We show that the second order field equations characterizing extremal solutions for spherically symmetric,
stationary black holes are in fact implied by a system of first order equations given in terms of a
prepotential $W$. This confirms and generalizes the results in \cite{anna}. Moreover  we prove that the
squared prepotential function shares the same properties of a c-function and that it interpolates between
$M^2_{ADM}$ and $M^2_{BR}$, the parameter of the near-horizon Bertotti-Robinson geometry. When the black
holes are solutions of extended supergravities we are able to find an explicit expression for the
prepotentials, valid at any radial distance from the horizon, which reproduces all the attractors of the
four dimensional $N>2$ theories. Far from the horizon, however, for $N$-even our ansatz poses a constraint
on one of the U-duality invariants for the non-BPS solutions with $Z\neq 0$.  We discuss a possible
extension of our considerations to the non extremal case.
\end{abstract}

 \vfill\eject
\section{Introduction} \label{intro}
Recent progress in the understanding of extremal non-BPS
black-hole solutions in extended supergravities (for a review on
black holes in supergravity see for example
\cite{review,reviewextr}) have revived the interest in the physics
of extremal black holes.

The peculiar feature of these solutions is the attractor mechanism
\cite{feka,fgk,gkk,k,tt,g,kss,k1,marrani,dst,dft}, according to
which the scalar ``hair'' of the black hole runs into a fixed
value on the horizon, independently of the boundary conditions at
spatial infinity. For static, spherically symmetric black holes,
the fixed values of the scalars at the horizon are determined in
terms of the quantized electric and magnetic charges
characterizing the solution,  as extrema of an effective potential
$V_{BH}$ \cite{gkk}. This induces to expect the radial dependence
of the scalar fields in each extremal solution to admit a
description in terms of a system of first order equations:
\begin{equation}
\dot \Phi^r \propto \partial^r W(\Phi)\,,
\end{equation}
which implies the second order field equations, provided $V_{BH}$
has a definite expression in terms of $W$ and its derivatives. In
this description, the attractor point $\dot\Phi=0$ is given by the
singular point $\partial_r W (\Phi_0)=0$ which is also an extremum
of $V_{BH}$. We will call $W$ the {\it prepotential} of the
extremal solution. This is indeed the case for BPS black holes,
whose associated first-order differential equations are implied by
the
  Killing-spinor equations. For the $N=2$ case one finds $W =|Z|$
  where $Z$ is the central charge of the supersymmetry algebra.
Therefore, the problem has raised of finding the  analogous first order differential equations, with the
associated prepotential, describing the non-BPS extremal solutions. Another motivation for developing a
first order formulation is that it provides a natural framework for defining a c-function associated with
 the radial flow of the fields in these solutions \cite{c-func}. In fact, as we will show, the squared
 prepotential is a viable candidate for such a function, sharing with it the monotonicity properties along a solution and the value taken at the horizon \footnote{We thank Sandip Trivedi for drawing this
 to our attention.}.

This problem was first addressed in \cite{anna} where, by exploiting the formal analogy  between  extremal
black holes and domain wall solutions,  explicit examples  of $W$ corresponding to certain $N=2$ non-BPS
extremal solutions were found.

It is the aim of the present paper to give a general form of  the prepotential  in extended supergravity
which will allow to reproduce the attractor behavior of  all the known extremal black-hole solutions for
$N\geq 3$. $W$ will be given as a function of the $U$-duality invariants of the theory built in terms of the
dressed charges. Different attractors will correspond to different choices of the coefficients in $W$. Let
us notice, however, that the general  ansatz we give for $W$ can be considered as a minimal one reproducing
correctly  all the attractor points of static extremal black-hole solutions in extended four dimensional
supergravity. However, if we consider the full black-hole solution outside the horizon, it turns out that
our ansatz requires a restriction, in the $N$-even cases, on the duality invariants characterizing the
non-BPS $Z\neq 0$ attractors. More precisely, for these solutions the above restriction amounts to fixing an
invariant overall phase of the complex dressed charges at radial infinity to the value it takes on the
horizon. We argue that a refined ansatz could relax this restriction.

The paper is organized as follows. In section two we recall  the main facts about static, spherically
symmetric black holes and introduce the prepotential $W$, proving, in the extremal case, that it is
monotonic. In section 3, which contains the main results of the paper, the general expression for $W$ in the
extremal case is given for $N\geq 3$ extended supergravity, and also some examples of $N=2$ solutions.
Section 4, which includes the concluding remarks, contains a speculative discussion where the issue of a
possible extension of the definition of the prepotential to the non extremal case is addressed. For a class
of non extremal black holes we show that a first order formulation in terms of a prepotential $W$ may exist
and we find the corresponding description in terms of first order differential equations. This generalizes
the results in \cite{weinberg} to the case of scalar-matter coupled gravity.

\section{Black holes as solutions to first order differential equations}

We will consider the class of  theories described by the bosonic
action \cite{feka}:
\begin{eqnarray}
 {\cal S}&=&\int\sqrt{-g}\,
d^4x\left(-\frac{1}{2}\,R+\Im{\cal N}_{\Lambda
\Gamma}F_{\mu\nu}^{\Lambda } F^{\Gamma |\mu\nu}+
\frac{1}{2\,\sqrt{-g}}\,\Re{\cal N}_{\Lambda \Gamma  }
\epsilon^{\mu\nu\rho\sigma}\, F_{\mu\nu}^{\Lambda } F^{\Gamma
}_{\rho\sigma}+\right.\nonumber\\
&+&\left.\frac 12 \,g_{rs}(\Phi) \partial_{\mu}
\Phi^{r}\partial^{\mu}\Phi^{s}\right)\,,\label{bosonicL}
\end{eqnarray}
where $R$ is the scalar curvature,  $\Phi^r$ are a set of scalar
fields and $F^\Lambda$ gauge field strengths. $g_{rs}(\Phi)$,
${\cal N}_{\Lambda\Sigma}(\Phi) $  ($r,s,\cdots =1,\cdots ,m$) are
matrices depending on the scalar fields.

The most general Ansatz for a spherically symmetric and stationary metric is
\begin{eqnarray}
ds^2&=&e^{2U}\,dt^2-e^{-2U}\,\left(\frac{c^4}{\sinh^4(c\tau)}\,d\tau^2+\frac{c^2}{\sinh^2(c\tau)}\,d\Omega^2\right)\,.\label{dstau}
\end{eqnarray}
The evolution coordinate $\tau$ is related to the radial coordinate $r$ by
the following relation:
\begin{eqnarray}
\left(\frac{dr}{d\tau}\right)^2=\frac{c^2}{\sinh^2(c\tau)}&=&(r-r_0)^2-c^2=(r-r^-)\,(r-r^+)\,.
\label{rtau}\end{eqnarray} $r_\pm$ being the radii of the two event horizons, with $r_+>r_-$. Here
$c\equiv 2ST$ is the extremality parameter of the solution, with $S$ the entropy and $T$ the temperature
of the black hole. In the extremal case $c\to 0$, eq. \eq{rtau} reduces to $\tau =-\frac{1}{r-r_H}$, where
$r_H$ denotes the radius of the horizon.

It is known \cite{gkk} that by eliminating the vector fields via
their equations of motion  this system may be reduced to the
following set of field equations for the metric function $U(\tau)$
and the scalar fields $\Phi^r(\tau)$ in terms of the evolution
parameter $\tau$:
 \begin{eqnarray}\frac{d^2 U}{d\tau^2}  &\equiv& \ddot U \,=\,
V_{\mbox{\small BH}}(\Phi,p,q)e^{2U}\,,\label{geoeq'}\\
 \frac{D^2
\Phi^r}{D\tau^2} &\equiv & \ddot\Phi^r +
\Gamma^r{}_{st}\dot\Phi^s\,\dot\Phi^t\,=\,
g^{rs}(\Phi)\,\frac{\partial V_{\mbox{\small
BH}}(\Phi,p,q)}{\partial \Phi^s} e^{2U} \,, \label{geoeq}
\end{eqnarray}
together with the constraint
\begin{equation}
\left(\frac{d U}{d\tau}\right)^2 +
\frac{1}{2}\,g_{rs}(\Phi)\,\frac{d \Phi^r}{d\tau}\frac{d
\Phi^s}{d\tau} - V_{\mbox{\small BH}}(\Phi,p,q)e^{2U}=c^2
\,,\label{bhconstr}
\end{equation}
where $V_{\mbox{\small BH}}(\Phi,p,q)$ is a function of the
scalars and of the electric and magnetic charges of the theory
defined by:
\begin{equation}
V_{\mbox{\small BH}}=-\frac{1}{2}\,Q^t\cM(\cN)Q\,, \label{geopot}
\end{equation}
and $Q$ is the symplectic vector of quantized  magnetic and
electric charges $Q^t=(p^\Lambda, q_\Lambda)$. $\cM(\cN)$ is the
symplectic matrix defined in terms of the gauge field-strengths
kinetic matrix $\cN_{\Lambda\Sigma}(\Phi)$:
\begin{eqnarray}
\cM (\cN) &=&   \pmatrix{  \Im \cN  +  \Re \cN \Im \cN^{-1}\Re \cN & - \Re \cN
\Im \cN^{-1}\cr- \Im \cN^{-1}\Re \cN & \Im \cN^{-1}\cr
}\,.
\label{m+}
\end{eqnarray}
The field equations \eq{geoeq} can be extracted from the
effective one-dimensional lagrangian:
\begin{equation}\cL_{eff}= \left(\frac{d U}{d\tau}\right)^2 +\frac{1}{2}\,
g_{rs}\frac{d \Phi^r}{d\tau}\frac{d \Phi^s}{d\tau} +
V_{\mbox{\small BH}}(\Phi,p,q)e^{2U} , \label{effect}
\end{equation}
 constrained with  equation \eq{bhconstr}.

We are going to show that the second order field equations \eq{geoeq'}, \eq{geoeq} can in fact be derived by a first order system, for a large class of extremal and non-extremal  black holes, by performing the following Ansatz \cite{anna}:
\begin{eqnarray}
\frac{dU}{d\tau}\,\equiv\,\dot{U}&=&e^U\,W(\Phi ,\tau)\,,\label{uw}
\end{eqnarray}
where $W$ is a function of the scalar fields (depending on  the
quantized charges and $\tau$)  and explicitly of $\tau$; the
derivative is performed with respect to the evolution parameter
$\tau$. We argue that the extremal case corresponds to
\begin{equation}
\partial_\tau W =0 \quad \Rightarrow \quad W=W(\Phi)\,,
\end{equation}
while for non extremal black holes, in those cases which admit a
first order description, an explicit dependence of $W$ on $\tau$
should be included.


\subsection{Extremal case} Let us   consider in detail the extremal
case $c=0$. In this case eq. \eq{uw} becomes
\begin{eqnarray}
\dot U = W(\Phi)\,e^U\,.
\label{udotextr}
\end{eqnarray}
Differentiating \eq{udotextr} with respect to $\tau$ gives
\begin{equation}
\ddot U=(\dot U)^2 +\dot W e^U\equiv W^2\,e^{2U}+\dot W e^U\,,
\label{u:}
\end{equation}
where
\begin{equation}
\dot W = \dot \Phi^r \partial_r W  \,.\label{Wdot}
\end{equation}
Comparing eq. (\ref{u:})  with \eq{geoeq'} we find the following
expression for $V_{BH}$:
\begin{equation}
V_{BH}=W^2 + e^{-U} \dot W.\label{pot1}
\end{equation}
Moreover from eqs. (\ref{geoeq'}) and (\ref{bhconstr}) one finds:
\begin{equation}
\ddot U-(\dot U)^2 = \frac 12 g_{rs} \dot \Phi^r \dot \Phi^s  =
\dot \Phi^r \partial_r W \, e^U\,, \label{constr2}
\end{equation}
while eq. (\ref{u:}) can be recast in the form:
\begin{equation}
\ddot U-(\dot U)^2 = \dot W\,e^U=\dot \Phi^r \partial_r W\,e^U\,.
\label{constr0}
\end{equation}
It follows that, for $\dot \Phi^r \neq 0$, is solved for
\footnote{To be precise, the most general solution to the constraint \eq{constr2}
 would be:
\begin{eqnarray}
\dot\Phi^r &=& 2\,e^U\, g^{rs}\,\partial_s W + \alpha^r (\Phi , \tau)\, ,
\end{eqnarray}
where
 $\alpha^r = P^{rs} h_s$ and
$P^{r}{}_s \equiv \left(\delta^{r}_s -\frac{ \dot \Phi^r \dot \Phi_s }{\dot \Phi^\ell \dot \Phi_\ell}\right)$ is a projector orthogonal to $\dot\Phi^r$.
In this more general case the effective potential would include one additional term:
\begin{eqnarray}
V_{BH}&=& W^2 + 2 \partial_r W \partial^r W - \frac 12  e^{-2U}
\alpha_r \alpha^r\,.
\end{eqnarray}
For $\alpha^r \neq 0$, however, the attractor condition at the horizon becomes
 \begin{eqnarray}
\lim_{\tau \to -\infty} \partial_r W = -\frac 12e^{-U} \alpha_r (\tau \to -\infty) \neq 0\,.
\end{eqnarray}
As we will see in the following (see eq. \eq{entropy}), such deformation is immaterial since it gives for the entropy the same value $V|_{extr}=W^2|_{extr}$ as for the case with $\alpha^r =0$.
It could instead play a role in more general situations, for example when considering black-holes out of extremality, where the effect of $\alpha^r$ would be to effectively deform the constant non-extremality parameter $c^2$ into a function $\mathcal{C}^2(\Phi,\tau) =   c^2 + \frac 12 \alpha_r \alpha^r$. }:
\begin{eqnarray}
\dot\Phi^r &=& 2\,e^U\, g^{rs}\,\partial_s W \label{phidot}\,.
\end{eqnarray}
Eq. \eq{phidot} is reminiscent of the BPS condition; for a given
$W$ it  relates the evolution of the scalar fields on the
corresponding configuration to the partial derivative of the
prepotential $W$ with respect to the scalar fields. Note in
particular that the fixed points for $\Phi^r$ are in direct
relation with the extrema of $W$. Together with  \eq{uw} and
\eq{Wdot}, \eq{phidot}  allows to express the field-equations in
terms of a first order system. Indeed, using  \eq{phidot}  the
effective potential reads \cite{anna}
\begin{eqnarray}
V_{BH}&=&W^2+2\,g^{rs}\,\partial_rW\,\partial_sW\,.\label{potextr}
\end{eqnarray}

By inserting eq.s \eq{phidot} and \eq{potextr} in the second order evolution equation for the scalars, eq. \eq{geoeq}, we find that it is {\em identically satisfied}.
 This shows that, as far as the scalar sector is concerned, the system of {\em second order} differential equations \eq{geoeq} is in fact  a {\em first order} system once expressed in terms of the prepotential $W$.
 Moreover,
given any explicit expression for $W$, also the space-time metric may be found as solution of a first-order equation \eq{udotextr}.
Furthermore, the effective potential \eq{potextr} is extremized for
\begin{eqnarray}
\frac{\partial V_{BH}}{\partial \Phi^r} \equiv \partial_r V_{BH}=
2 \, \partial_s W \left(W
\delta^s_r+2g^{s\ell}\nabla_r\partial_\ell
W\right)=0\,,\label{min0}
\end{eqnarray}
that is the fixed points for the scalars (corresponding to extrema
of $W$) are also extrema of the potential $V_{BH}$. Since the
black-hole horizon is identified as the fixed point of the
scalars, in this formulation it is directly related to the extrema
of $W$ (that are in particular also extrema for $V_{BH}$. The BH
entropy then reads, in terms of $W$, as
\begin{eqnarray}
S_{BH} =V_{BH}|_{extr}=W^2|_{extr}\label{entropy}
\end{eqnarray}
Furthermore, let us observe that from the evolution equations above, together with the boundary condition on
$U$ at spatial infinity ($U(\tau =0)=0$), it is easy to deduce that $W$ and $W^2$ are monotonic functions,
both decreasing along the evolution from spatial infinity towards the horizon. Indeed, if we define the
function $b(\tau)= -\frac{1}{\tau}\, e^{-U}$, as in \cite{tt}, the conditions of regularity of the solution
at the horizon and of flatness of space-time at radial infinity, imply the following limiting behaviors:
\begin{eqnarray}
\tau \to -\infty &:& b(\tau) \longrightarrow r_{H} + \mathcal{O}(\tau^{-1}) \\
\tau \to 0^- &:& b(\tau) \longrightarrow -\frac{1}{\tau} + M_{ADM} + \mathcal{O}(\tau) \,.
\end{eqnarray}
Since $e^{-U} = -\tau b$, using the first order equations we may write
\begin{eqnarray}
W= -\frac{d}{d\tau} e^{-U} = \frac{d}{d\tau} (\tau b) = b+\tau \dot b\,,
\end{eqnarray}
which implies the following asymptotic limits for $W$:
\begin{eqnarray}
\lim_{\tau\rightarrow-\infty} W&=&r_H=M_{BR}\,,\nonumber\\
\lim_{\tau\rightarrow 0^-} W&=&M_{ADM}\ge M_{BR}\,,
\end{eqnarray}
where $M_{BR}$ denotes the Bertotti-Robinson mass parameter associated with the near-horizon geometry. Let
us now show that $W$ is monotonic:
\begin{eqnarray}
\frac{d W}{d\tau}&=&\left(\ddot{U}-(\dot U)^2\right)\,e^{-U}=\left(V-W^2\right)\,e^{U}=2\,g^{rs}\,\partial_r
W\partial_sW\,e^{U}\ge 0\,,
\end{eqnarray}
where we have used eq. (\ref{potextr}) and  the first order equations. We conclude that $W$ is a positive
monotonic function decreasing from the value $M_{ADM}$ at radial infinity, towards the value $M_{BR}\le
M_{ADM}$ at the horizon.
 In \cite{c-func} it was
shown that for static, spherically symmetric black holes a monotone function $A(r)$ always exists such that
$A(r_H)=S_{BH}$, with $A(r)$ decreasing towards the horizon. For the extremal case, $W^2$ then appears as
the appropriate quantity to play the role of the c-function.

\section{The prepotential for extremal solutions of extended supergravity}
\label{extremalW}

It is our purpose to show that when the extremal black hole is a
solution of an $N>2$ extended  supersymmetric theory, where the
scalar manifold is a coset $G/H$, it is possible to find a general
expression for $W$ which reproduces all the known results
concerning the BPS and non-BPS attractor points of the theory (see
\cite{reviewextr} for a review collecting all the solutions in
four dimensions). Our results may be extended to the $N=2$ case
when the special geometry is described by homogeneous spaces. For
more general models a case by case inspection is necessary. Some
$N=2$ examples have been given in \cite{anna}.

We propose the following:
\begin{eqnarray}
W =\sum_M \alpha^M e_M \,.\label{Wextremal}
\end{eqnarray}
where $e_M\in \IR$ are
 related to the invariants of the isotropy subgroup $H$ built in terms of the complex central and matter charges.
Such invariants indeed can be expressed in terms of the skew eigenvalues of the matrix of central charges $Z_{AB}$ and of the norm of the matter charge vectors $Z_I$ that we collectively call $\{e_M\}$.

The real coefficients $\alpha^M$ can be computed by requiring that the potential \eq{potextr},
with $W$ given by \eq{Wextremal}, reproduces  the general form taken by the effective scalar potential for
 any extended supersymmetric theory:
\begin{eqnarray}
V_{BH} = \frac 12 Z_{AB} \bar Z^{AB} + Z_I \bar Z^I
\,.\label{Vsusy}
\end{eqnarray}
Here and in the following $A,B$ are ${\rm SU}(N)$ R--symmetry
group  indices while the indices $I,J$ label the fundamental
representation of the matter group when present (namely ${\rm
U}(3)$ for $N=3$ and ${\rm SO}(6)$ in the $N=4$ case). Since
\eq{potextr} involves the gradient of the prepotential $W$, the
evaluation of $V_{BH}$ requires the knowledge of the differential
relations among central and matter charges, for which we refer to
\cite{urevisited,reviewextr}. In general, since the equations in
the $\alpha^M $ are quadratic, their sign is not fixed in
principle, but it can be fixed by requiring that the prepotential
$W$ is extremized on the black-hole horizon $\partial_r
W|{hor}=0$.

It was observed in \cite{anna} that for any $V_{BH}$ it should
exist a multiple choice of $W$. Here, using the ansatz
\eq{Wextremal} we give  the explicit expression and the precise
number of independent prepotentials for any given extended theory.
Indeed, as we are going to show by a case by case analysis, there
are in general up to three independent choices of $\{\alpha^M\}$
all reproducing the same $V_{BH}$. The various independent
solutions for $W$ will reproduce the known different BPS and
non-BPS solutions for any given theory.
 Any independent choice of $\{\alpha^M\}$ would then parametrize a different black-hole solution.

We may then adopt  two equivalent points of view to find the
different extremal black-hole attractors: either we study the
extrema of $V_{BH}$, or alternatively we consider the possible
inequivalent choices of $W$ compatible with the expression
\eq{Vsusy} for $V_{BH}$.

We first analyze the $N$-odd cases, which are easier because the central and matter charges in normal form  can  all be made real. The cases with $N$ even, which in general also include a solution with complex charges (corresponding to a negative fourth-order invariant of the duality group $G$) will be analyzed afterwords.


\subsection{ The $N$-odd cases}\label{oddcases}
Since these are the simplest cases, we shall describe the
calculations in detail. In the $N$--even cases the relevant
results will be given.
\subsubsection{The $N=3$ case}\label{n=3}
 In the  $N=3$ theory the scalar manifold is $U(3,n)/[U(3)\times U(n)]$ and the central charge matrix $Z_{AB} =-Z_{BA}$, $A=1,2,3$, and matter charges $Z_I$, $I=1,\cdots ,n$;  the central and matter charges obey the differential relations
 \begin{eqnarray}
\left\{\matrix{ \nabla Z_{AB} &=& P_{IAB} \bar Z^I \cr
 \nabla Z_I &=&\frac 12 P_{IAB} \bar Z^{AB} \,,}\right. \label{diffrel3}
 \end{eqnarray}
where $P_{IAB} = P_{IAB,i}dz^i$ ($i=1,\cdots 3n$) is the holomorphic vielbein of $U(3,n)/[U(3)\times U(n)]$,
$\nabla$ denotes the ${\rm U}(1)$--K\"ahler covariant and $H$--covariant derivative, where $H$ is the
isotropy group of the symmetric spaces $G/H$ repres4nting the scalar manifold of the various $N\ge 3$
theories (see \cite{reviewextr,urevisited}). By a $U(3)$ rotation it is always possible to put $Z_{AB}$ in
normal form
 \begin{eqnarray}
 Z_{AB}= e \pmatrix{0&1&0\cr
 -1 &0&0\cr
 0&0&0}\, \qquad e\in \IR
 \end{eqnarray}
 while by a $U(n)$ rotation the vector $Z_I$ may by chosen to be real and pointing in a given direction, say
 \begin{eqnarray}
 Z_I =\rho \delta_I^1\quad \rho\,\in \,\IR .
 \end{eqnarray}
 We then propose the following general expression for $W$:
 \begin{eqnarray}
 W &=& a \,e + b \,\rho \nonumber \\
 &=& a \sqrt{\frac 12 Z_{AB} \bar Z^{AB}}+ b \sqrt{ Z_{I} \bar Z^{I}}\,.\label{W3}
 \end{eqnarray}
 From the relation (\ref{potextr}) one obtains:
 \begin{eqnarray}
 V = \left(a^2+b^2\right)\left( e^2 + \rho^2\right) + 4ab \,e \rho.
 \end{eqnarray}
In order to reproduce the general result \eq{Vsusy}, that written in normal form takes the
form
 \begin{eqnarray}
 V= e^2 + \rho^2 \,,
 \end{eqnarray}
 we must have:
 \begin{eqnarray}
 \cases{a^2 + b^2 =1\,, \cr
 ab = 0\,.} \label{cond3}
 \end{eqnarray}
 There are two different  solutions to the system \eq{cond3}, namely
\begin{enumerate}
\item{ $a=1$, $b=0$, implying
\begin{eqnarray}
W_{(1)}=e = \frac 12 \sqrt{Z_{AB}\bar Z^{AB}}\end{eqnarray} }
\item{ $a=0$ and  $b=1$, which  imply
\begin{eqnarray}
W_{(2)}=\rho =  \sqrt{Z_{I}\bar Z^{I}}\,.\end{eqnarray}}
\end{enumerate}
Note that these two choices reproduce precisely the two
independent solutions for the extremization of the black-hole
solutions of $N=3$ supergravity \cite{reviewextr}.  The former
solution, which implies:
\begin{eqnarray}
\nabla_{i} W_{(1)} =\frac 1{4\sqrt{{\frac12} Z_{AB}\bar Z^{AB}
}
}
\bar Z^I \bar Z^{AB} P_{IAB,i}
\end{eqnarray}
is extremized for $Z_I=0$, $Z_{AB}\neq 0$ and corresponds to the
BPS solution, with entropy $S_{(1)}= W_{(1)}^2|_{extr}=\frac 12
|Z_{AB}|^2$. The second one gives
\begin{eqnarray}
\nabla_{i} W_{(2)} =\frac 1{4\sqrt{ Z_{I}\bar Z^I}}\bar Z^I \bar Z^{AB} P_{IAB,i}
\end{eqnarray}
and is extremized for
$Z_{AB}=0$, $Z_I \neq 0$  corresponding to the non-BPS solution, with entropy $S_{(2)}= W_{(2)}^2|_{extr}= |Z_{I}|^2$.

Since we know that the potential \eq{Vsusy} has two minima for the
$N=3$ theory, $W_{(1)}$ and $W_{(2)}$
 exhaust the possible minima of the general potential for this theory.  This in particular
 implies that eq. \eq{min0} cannot have further solutions coming from the vanishing of the second factor, as
 it can be easily shown by an explicit calculation.

\subsubsection{The prepotential for the $N=5$ case}\label{n=5}
In this case there are no matter multiplets and  the scalar manifold is the K\"ahler manifold
$SU(1,5)/U(5)$, spanned by the holomorphic vielbein $P_{ABCD}=P_{ABCD,i}\,dz^i=\epsilon_{ABCDE} P^E$ (with
$A,i=,\cdots 5$) and its complex conjugate $P^{ABCD}=P^{ABCD}_{,\ibar}\,d\bar z^\ibar$. The central charges
$Z_{AB} =-Z_{BA}$ obey the differential relations:
\begin{eqnarray}
\nabla Z_{AB} = \frac 12 P_{ABCD} \bar Z^{CD}\,.\label{diffrel5}
\end{eqnarray}
Via a $U(5)$ rotation they may be put in the normal form:
\begin{eqnarray}
Z_{AB}= \pmatrix{0&e_1 &0&0&0\cr
-e_1&0&0&0&0\cr
0&0&0&e_2&0\cr
0&0&-e_2&0&0\cr
0&0&0&0&0}
\end{eqnarray}
in terms of the two real (non negative) proper-values $e_1$ and $e_2$, which are related to the two $U(5)$ invariants
 \begin{eqnarray}
\left\{\matrix{I_1 &\equiv& \frac 12 Z_{AB} \bar Z^{AB} =(e_1)^2 + (e_2)^2 \cr
I_2 &\equiv&\frac 12 Z_{AB} \bar Z^{BC}Z_{CD} \bar Z^{DA}=(e_1)^4 + (e_2)^4}\right.
\end{eqnarray}
by the inverse relation
 \begin{eqnarray}
\left\{\matrix{e_1 & =&\sqrt{\frac12 \left[I_1 + \sqrt{2I_2-I_1^2}\right] }\cr
e_2 & =&\sqrt{\frac12 \left[I_1 - \sqrt{2I_2-I_1^2}\right]}}\right..\label{pv5}
\end{eqnarray}
According to equation \eq{Wextremal}, we then propose for the prepotential $W$ the form
\begin{eqnarray}
W=a_1 e_1 + a_2 e_2 \,.
\end{eqnarray}
Writing (\ref{diffrel5}) in normal form, that is
\begin{eqnarray}
\nabla_ie_1&=&P_{,i}\,e_2\,,\nonumber\\
\nabla_ie_2&=&P_{,i}\,e_1\,,
\end{eqnarray}
 Its holomorphic gradient in normal form is
\begin{eqnarray}
\partial_i W=\frac 12 P_{,i} (a_1 e_2 + a_2 e_1) \,.\label{dW5}
\end{eqnarray}
where $P_{,i} =P_{1234,i}$ is the component of the holomorphic scalar vielbein which appears
in \eq{diffrel5} when the central charge is  in its normal form. Evaluating the potential
using \eq{diffrel5}  gives, for the black-hole potential,
\begin{eqnarray}
V_{BH}= (a_1^2 + a_2^2)\left[(e_1)^2 + (e_2)^2 \right]+ 4a_1a_2
e_1 e_2 \label{pot5}\,.
\end{eqnarray}
This reproduces the result for the black-hole potential of supersymmetric theories, $V=I_1$,
for
 \begin{eqnarray}
\left\{\matrix{a_1^2 +a_2^2&=&1\cr
a_1a_2&=&0}\right.
\end{eqnarray}
This system has, essentially, only one independent solution that, with our choice  \eq{pv5} of proper-values
(which implies $e_1\geq e_2$) is
\begin{eqnarray}
a_1=1\,,\quad a_2=0
\end{eqnarray}
giving
$W=e_1$, which is extremized for $\nabla_i W = e_2 =0$.
It is a BPS solution.

Note that the extremization of the scalar potential \eq{pot5} gives
\begin{eqnarray}
\partial_i V_{BH} = 2 \partial_k W \left(\delta_i^k W + 2 g^{k\jbar}\nabla_i \partial_\jbar W\right)=0,
\end{eqnarray}
since, from \eq{dW5}, $\nabla_i\partial_k W=0$. It may be easily
shown by explicit calculation that there are no other solutions to
$\partial_i V_{BH}=0$ besides $\partial_i W=0$
from which we conclude that also in this case the extrema of $W$
give all the extrema of $V_{BH}$.

\subsection{The $N$-even cases}\label{evencases}
In the cases with $N=3$ and $N=5$ supercharges the central and
matter charges  $Z_M\equiv\{Z_{AB},\,Z_I\}$ in normal form may be
chosen all real and non negative. On the other hand, in the
$N$-even cases the normal form of the $Z_M$ contains in general
an unfixed overall phase.  Since our choice \eq{Wextremal} of the
prepotential is given only in terms of the moduli of the charges,
the solution for the coefficents $\alpha^M$ in \eq{Wextremal}
which reconstruct the effective potential $V_{BH}$ implies in
particular that the phase must be fixed at certain values all over
the moduli space. This value is actually the one corresponding to
the attractor condition at the horizon.

Our general  ansatz \eq{Wextremal} can then be considered as the minimal one reproducing correctly all the
attractor points of static extremal black-hole solutions in extended four dimensional supergravity. We argue
that a refined ansatz could relax the fixing of the phase before extremization.
\subsubsection{The prepotential for the $N=4$ attractors}\label{n=4}

In this case the scalar manifold  is the coset space
\begin{equation}
G/H=\frac{SU(1,1)}{U(1)} \times \frac{SO(6,n)}{SO(6)\times SO(n)}\label{n=4coset}
\end{equation}
and the relations among central and matter charges are:
\begin{eqnarray}
 \left\{\matrix{ D Z_{AB} &=&   \bar Z^{I}  P_{AB I}+  \frac{1}{2} \bar Z^{CD} \epsilon_{ABCD}\, P\,, \cr
D  Z_{I} &=& \frac{1}{2}  \bar Z^{AB}  P_{AB I} +
\bar Z_I  \,\bar P\,.\hfill}\right.\label{n=4rel}
\end{eqnarray}
We recall that for this theory the vielbein $P_{AB I}$ satisfies the reality condition
$\bar P^{AB I} \equiv (P_{AB I})^\star = \frac 12 \epsilon^{ABCD}  P^I_{CD }$.

Using the $\rm U(1)\times\rm SO(6)\sim \rm U(4)$ symmetry of the theory we can bring the
central charges into their normal form \cite{zumino}
\begin{eqnarray}
Z_{AB}= \pmatrix{0&Z_1 &0&0\cr
-Z_1&0&0&0\cr
0&0&0&Z_2\cr
0&0&-Z_2&0}
\end{eqnarray}
where  the graviphoton skew-eigenvalues $Z_1,\,Z_2 $ can  be chosen real and non-negative, thus  coinciding
with their modulus $Z_{1,2} =|Z_{1,2}|=e_{1,2}$. Further,  using an $\rm SO(n)$ transformation  it is also
possible to reduce the vector of matter charges in such a way that only one real and one complex matter
charge are different from zero. Let us call them $Z_I =\rho_I e^{\ii \theta_I}$, $I=1,2$ (with the proviso
that one of the phases may be always put to zero).

 We consider  a prepotential $W$ of
the form:
\begin{eqnarray}
W&=&a_1\,e_1+a_2\,e_2+\sum_{I=1}^2 b_I\,\rho_I\,.
\end{eqnarray}
It encodes  the $H$ and $G$ (moduli-dependent) invariants:
 \begin{eqnarray}
\left\{\matrix{I_1 &\equiv& \frac 12 Z_{AB} \bar Z^{AB} =(e_1)^2 + (e_2)^2\,,\hfill \cr I_2
&\equiv&\frac 12 Z_{AB} \bar Z^{BC}Z_{CD} \bar Z^{DA}=(e_1)^4 + (e_2)^4\,,\hfill\cr
I_3&\equiv& Z_I \bar Z_I = \rho_I \rho_I\,, \hfill\cr I_4 &\equiv & \Re \left(Z_I
Z_I\right)\,. \hfill}\right.
\end{eqnarray}

 The potential is related
to $W$ by eq. (\ref{potextr}).
 In order to compute the derivatives of $W$ we rewrite, as usual, the differential relations in normal form,
   where $P =P_{,i }dz^i$ is the K\"ahlerian
 vielbein of $\rm SU(1,1)/U(1)$ while $P_{12 I} \equiv P_I$ ($P_{34 I} = \bar P_{12 I}$)
 are the components of the (non K\"ahlerian) vielbein $\rm SO(6,n)/SO(6)\times SO(n)$:
\begin{eqnarray}
 \left\{\matrix{ \nabla Z_{1} &=&   \bar Z^{I}  P_{1 I}+   \bar Z_{2} \, P\,, \hfill\cr
\nabla Z_{2} &=&   \bar Z^{I}  \bar P_{1 I}+   \bar Z_{1} \, P\,, \hfill\cr \nabla  Z_{I} &=&
\bar Z_1  P_{1 I}+ \bar Z_2 \bar P_{1I} + \bar Z_I  \,\bar P\,.\hfill}\right.\label{n=4relnf}
\end{eqnarray}
 We then find:
 \begin{eqnarray}
\nabla_i W&=&P_{,i}\,A+P_{1I,i}\,B_I + \bar P_{1I,i}\,\bar B_I  \,,\nonumber \\
A&=&\frac{1}{2}\left[(a_1\,e_2+a_2\,e_1)+\sum_I
b_I\rho_I \,e^{2\,\ii\,\theta_I}\right]\,,\nonumber\\
B_I&=&\frac{1}{2}\left[(a_1\rho_I+b_I e_1)\,e^{-\,\ii\,\theta_I}+
(a_2\rho_I+b_I e_2)\,e^{\,\ii\,\theta_I}\right]\,.\label{dw4}
 \end{eqnarray}
In terms of the above quantities  the potential reads:
\begin{eqnarray}
 W^2+2\,g^{rs}\partial_rW\partial_sW&=&\left(||a||^2+||b||^2\right)(|Z_1|^2+|Z_2|^2+\sum_I \rho_I^2
)+\nonumber\\
 &&+\sum_{I\neq J}\left(b_I^2-b_J^2+2\,a_1\,a_2\,\cos(2\,\theta_I)\right)\,\rho_I^2+
 \nonumber\\&&+
2\,|Z_1||Z_2|\left(2\,a_1\,a_2+\sum_I b_I^2\,\,\cos(2\,\theta_I)\right) +\nonumber\\
\hskip -5mm && +4\,\sum_I|Z_1|\rho_I \,b_I\left(a_1+a_2\,\cos(2\,\theta_I)\right)
+\nonumber\\&&
 +4\,\sum_I|Z_2|\rho_I
\,b_I\left(a_2+a_1\,\cos(2\,\theta_I)\right)
+\nonumber\\
\hskip -5mm &&
+2\,b_1\,b_2\,\rho_1\rho_2\,\left(\cos{2\,(\theta_1-\theta_2)}+1\right)\,,\label{Vmess}
\end{eqnarray}
where we have defined: $||a||^2=a_1^2+a_2^2,\,||b||^2=\sum_Ib_I^2$. Since the potential in
terms of the central charges in the normal form reads
\begin{eqnarray} V&=& |Z_1|^2+|Z_2|^2+\sum_I \rho_I^2\,.
\end{eqnarray}
the coefficients $a_k,\,b_I$ and the phases have to be chosen in such a way that the cross
terms in the central charges in (\ref{Vmess}) vanish and the coefficients of the square norm
of the central charges be equal to one. This implies that:
\begin{eqnarray}
\left\{\matrix{&&||a||^2+||b||^2=1\,,\hfill\cr && b_I^2 -b_J^2 +2\,a_1\,a_2\,\cos(2\,\theta_I)=0\,,\quad
\,\,\,\,\,\forall I\neq J\,,\hfill\cr &&2\,a_1\,a_2+\sum_I b_I^2\,\,\cos(2\,\theta_I)=0\,,\hfill\cr &&
b_I\left(a_1+a_2\,\cos(2\,\theta_I)\right)=0\quad \forall I\,,\hfill\cr &&
b_I\left(a_2+a_1\,\cos(2\,\theta_I)\right)=0\quad \forall I\,,\hfill\cr &&
b_1\,b_2\,\left(\cos{2\,(\theta_1-\theta_2)}+1\right)=0\,.\quad \hfill}\right.\label{Vcond}
\end{eqnarray}
In the following we choose normal form of the matter charges so
that $\theta_2=0$.
 The above conditions can be explicitly solved giving three independent solutions for the coefficients.
 Consequently we have three different prepotentials, each characterizing a different attractor
 solution. The inequivalent solutions  are:
 \begin{enumerate}
 \item{$a_1=1,\,a_2=0,\,b_I=0$ or $a_1=0,\,a_2=1,\,b_I=0,\,\forall I$.

The prepotential reads:
\begin{eqnarray}
W_{(1)}&=&W_{BPS}= e_1\,. \end{eqnarray} The attractor condition for this solution gives
indeed, from \eq{dw4}:
\begin{eqnarray}
\partial_iW_{(1)}=0 \Rightarrow e_2 = \rho_I =0.
\end{eqnarray}
This corresponds to the BPS attractor, with entropy
$S_1=W_{1|extr}^2=e_1^2$.}
\item{
 $a_1=a_2=\frac{1}{\sqrt{2}} b_1= \frac 12,\quad
b_2=0,\quad\theta_1=\frac{ \pi}{2}$.
The complete choice of the
prepotential, fixed by the attractor condition  at the horizon,
gives:
\begin{eqnarray}
W_{(2)}&=& \frac12(e_1 + e_2 + \sqrt 2 \rho_1)\,,
\end{eqnarray}
which is indeed extremized for:
\begin{equation}
e_1=e_2= e\,; \quad Z_{I=1}= \sqrt 2 \ii e\,;\,\,\,\,\,Z_{I=2}=0\,.
\end{equation}
The entropy is given by $S_2=W_{2|extr}^2=4\,e^2$. }
\item
{$a_1=a_2=0,\,\,\,\,b_1=b_2=\frac{1}{\sqrt{2}}\,\,\,,\,\,\theta_1=\frac{
\pi}{2}$.
 The prepotential reads
\begin{eqnarray}
W_{(3)}&=&\frac{1}{\sqrt{2}}\,(\,\rho_1 + \rho_2)\,.
\end{eqnarray}
The extremum condition for this solution is
\begin{eqnarray} e_1=e_2=0\,\quad Z_{I=2}= \ii Z_{I=1}=\rho\,.
\end{eqnarray}
The entropy is $S_3=W_{3|extr}^2=\rho^2$.}
\end{enumerate}

\subsubsection{The prepotential for the $N=6$ theory}\label{n=6}
In this case the scalar manifold is $SO^*(12)/U(6)$, spanned by the holomorphic vielbein
$P_{ABCD}=P_{ABCD,i}\,dz^i=\frac 12 \epsilon_{ABCDEF} P^{EF}$ (with $A=1,\cdots 6$,
$i=1,\cdots 15$) and its complex conjugate $P^{ABCD}=P^{ABCD,\ibar}\,d\bar z^\ibar$. The
central charges of this theory are split into an antisymmetric matrix $Z_{AB} =-Z_{BA}$ and a
singlet $X$. They obey the differential relations
\begin{eqnarray}
\left\{\matrix{\nabla Z_{AB} &=& \frac 12 P_{ABCD} \bar Z^{CD} + \frac 1{4!}\epsilon_{ABCDEF} \bar P^{CDEF} \bar X\cr
\nabla X &=& \frac 1{2!4!}\epsilon_{ABCDEF} \bar P^{CDEF} \bar Z^{AB}}\right.  \,.\label{diffrel6}
\end{eqnarray}
The singlet complex charge $X$ may be parametrized as $X=\rho\, {\mathrm{e}}^{\ii\alpha}$  (with $\rho\in \IR_+$,  $\alpha \in \IR$). On the other hand the antisymmetric matrix $Z_{AB}$ may  be put in the normal form  via a $U(6)$ rotation:
\begin{eqnarray}
Z_{AB}= \pmatrix{0&Z_1 &0&0&0&0\cr
-Z_1&0&0&0&0&0\cr
0&0&0&Z_2&0&0\cr
0&0&-Z_2&0&0&0\cr
0&0&0&0&0&Z_3\cr
0&0&0&0&-Z_3&0}
\end{eqnarray}
in terms of the three proper-values $Z_1$, $Z_2$, $Z_3$. In the normal form they may indeed be chosen real and non negative $Z_{\alpha}=|Z_\alpha|\equiv e_\alpha$ ($\alpha=1,2,3$).

 The four parameters $e_\alpha, \rho $  are related to the four $U(6)$ invariants:
 \begin{eqnarray}
\left\{\matrix{
I_1 &\equiv& \frac 12 Z_{AB} \bar Z^{AB} =(e_1)^2 + (e_2)^2+ (e_3)^2\hfill \cr
I_2 &\equiv& X\bar X = \rho^2\hfill\cr
I_3 &\equiv&\frac 12 Z_{AB} \bar Z^{BC}Z_{CD} \bar
Z^{DA}=(e_1)^4 + (e_2)^4+ (e_3)^4\hfill\cr
 I_4&\equiv& - \frac 12 Z_{AB} \bar Z^{BC}Z_{CD} \bar
Z^{DE}Z_{EF} \bar Z^{FA}=(e_1)^6 + (e_2)^6+ (e_3)^6\,.\hfill}\right.\label{i6}
\end{eqnarray}

Writing the differential relations among the dressed charges \eq{diffrel6}  in normal form, we find a simple
expression for the holomorphic derivatives of the skew-eigenvalues $e_\alpha$, namely:
\begin{eqnarray}
\left\{\matrix{\nabla_i e_1 &=& \frac12 \left(P_{1,i}e_2 + P_{2,i}\rho \e^{\ii\alpha} + P_{3,i} e_3\right)\cr
\nabla_i e_2 &=& \frac12 \left(P_{1,i}e_1 +  P_{2,i} e_3+P_{3,i}\rho \e^{\ii\alpha} \right)\cr
\nabla_i e_3 &=& \frac12 \left(P_{1,i}\rho \e^{\ii\alpha} + P_{2,i} e_2+ P_{3,i} e_1\right)\cr
\nabla_i \rho &=& \frac 12 \e^{\ii\alpha}\left(P_{1,i}e_3 + P_{2,i}e_1 + P_{3,i} e_2\right)\cr
\nabla_i \alpha &=& \frac \ii{2\rho} \e^{\ii\alpha}\left(P_{1,i}e_3 + P_{2,i}e_1 + P_{3,i} e_2\right)
}\right. \label{de6}
\end{eqnarray}
where $P_{1,i}=P_{1234,i}$, $P_{2,i}=P_{3456,i}$, $P_{3,i}=P_{1256,i}$ are the components of
the scalar vielbein appearing in \eq{de6} when  the central charge is written in normal form.
We then propose, for the prepotential $W$, the form:
\begin{eqnarray}
W=a_1 e_1 + a_2 e_2 +a_3 e_3 + b \rho \,,\label{W6}
\end{eqnarray}
giving, for its holomorphic gradient in normal form:
\begin{eqnarray}
\nabla_i W &=& \frac12\Bigl\{P_{1,i}\left[a_1 e_2 +a_2 e_1
 + {\mathrm{e}}^{\ii\alpha}(a_3 \rho +b e_3)\right]+\nonumber\\
&&+P_{2,i}\left[a_2 e_3 +a_3 e_2 + {\mathrm{e}}^{\ii\alpha}(a_1 \rho +b e_1)\right]+\nonumber\\
&&+P_{3,i}\left[a_1 e_3 +a_3 e_1 + {\mathrm{e}}^{\ii\alpha}(a_2 \rho +b e_2)\right]\Bigr\}\,.\label{dW6}
\end{eqnarray}
Using eqs. \eq{de6} and  eq. \eq{W6},  the right-hand side of \eq{potextr} takes the following form:
\begin{eqnarray}
W^2+4\,g^{i\bar\jmath}\partial_iW\partial_{\bar\jmath}W&=& (a_1^2 + a_2^2 +a_3^2 + b^2)\left(e_1^2 +e_2^2+e_3^2 +b^2\right) +\nonumber\\
&&+ 2e_1e_2 (a_1a_2 + a_3 b \cos\alpha )+ 2 e_3 \rho (a_1a_2 \cos\alpha + a_3 b )+ \nonumber\\
&&+ 2e_2e_3 (a_2a_3 + a_1 b \cos\alpha )+ 2 e_1 \rho (a_2a_3 \cos\alpha + a_1 b )+ \nonumber\\
&&+ 2e_3e_1 (a_3a_1 + a_2 b \cos\alpha )+ 2 e_2 \rho (a_3a_1 \cos\alpha + a_2 b
)\end{eqnarray} which reproduces the result for the black-hole potential of supersymmetric
theories, \begin{equation} V=I_1+I_2=e_1^2 + e_2^2 + e_3^2 + \rho^2 \, \end{equation} if:
\begin{eqnarray}
\left\{\matrix{a_1^2 + a_2^2 +a_3^2 + b^2&=&1\hfill\cr a_1a_2 +
a_3 b \cos\alpha &=&0\hfill\cr a_1a_2  \cos\alpha+ a_3 b
&=&0\hfill\cr a_1a_3 + a_2 b \cos\alpha &=&0\hfill\cr
a_1a_3\cos\alpha + a_2 b   &=&0\hfill\cr a_2a_3 + a_1 b \cos\alpha
&=&0\hfill\cr a_2a_3\cos\alpha  + a_1 b  &=&0\,.\hfill}\right.
\label{sys6}
\end{eqnarray}
This system, together with the requirement of the existence of an
attractor at the horizon, allows three inequivalent solutions, all
of which requiring the phase of the singlet charge to be fixed,
when the system is in normal form, by $\cos^2\alpha =1$.
\begin{enumerate}
\item $a_1=1$, $a_2=a_3 =b=0$ (or, equivalently, for $a_1 \leftrightarrow a_2 \leftrightarrow a_3$).

In this case the prepotential encoding the solution has the form:
\begin{eqnarray}
W_{(1)}&=& e_1\,,
\end{eqnarray}
which is extremized (see \eq{dW6}), for $e_1=e$, $e_2=e_3=\rho=0$. The Bekenstein--Hawking
entropy is:
\begin{eqnarray}
V_{(1)}|_{extr}= e^2\,.
\end{eqnarray}
This is the BPS solution. \item $a_1=a_2=a_3=0$, $b=1$.

In this case the prepotential is:
\begin{eqnarray}
W_{(2)}&=& \rho =\sqrt{X\bar X}\,.
\end{eqnarray}
It is extremized for $e_1=e_2=e_3 =0$, and it is a non-BPS solution. The corresponding entropy
is:
\begin{eqnarray}
V_{(2)}|_{extr}=\rho^2\,.
\end{eqnarray}
\item $a_1=a_2=a_3=b= \frac 12$.

This solution requires that  the phase of the singlet
 charge $X$ be fixed to $\alpha=\pi$: $X=-\rho$.
We then have:
\begin{eqnarray}
\left\{\matrix{W_{(3)}&=& \frac12 (e_1+e_2 + e_3+ \rho) \cr \alpha&=&\pi \,. \hfill}\right.
\end{eqnarray}
$W$ is extremized  for: $e_1=e_2=e_3 =e$, $X=-e$.
 In this case the entropy is
\begin{eqnarray}
V_{(3)}|_{extr}=4 e^2
\end{eqnarray}
This is also a non-BPS extremal solution.
\end{enumerate}
Note that since the bosonic sector of the $N=6$ theory also describes an $N=2$ model, the
three solutions above are also solutions of the equivalent $N=2$ model based on the coset $\rm
SO^*(12)/U(6)$. In this case, however, the singlet charge $X$ is the central charge
corresponding to the $N=2$ graviphoton, while the $Z_{AB}$ are matter charges, so that the
first two solutions are interchanged in the $N=2$ version: the first one is non-BPS while the
second one is the BPS solution.

\subsubsection{The prepotential for the $N=8$ theory}\label{n=8}

The scalar manifold of the $N=8$ theory is $\rm E_{7(-7)}/SU(8)$. It is not a K\"ahler manifold, and it is
spanned by the vielbein $P_{ABCD}=\frac 1{4!}\epsilon_{ABCDEFGH}\bar P^{EFGH}$ (with $A=1,\cdots 8$). The
central charges of this theory belong to an antisymmetric matrix $Z_{AB} =-Z_{BA}$. They obey the
differential relations:
\begin{eqnarray}
\nabla Z_{AB} &=& \frac 12 P_{ABCD} \bar Z^{CD}   \,.\label{diffrel8}
\end{eqnarray}
Since for this theory, differently from the other four dimensional cases,  the holonomy group
does not contain a $U(1)$ factor, when the antisymmetric matrix $Z_{AB}$ is put in   normal
form,  via an $SU(8)$ rotation, it still depends on an overall phase. Therefore we can write:
\begin{eqnarray}
Z_{AB}= \e^{\ii\frac\alpha 4}\pmatrix{e_1 &0&0&0\cr 0&e_2&0&0\cr 0&0&e_3&0\cr
0&0&0&e_4}\otimes \pmatrix{0&1\cr-1&0}\,.
\end{eqnarray}
in terms of the four real (non negative) skew-eigenvalues $e_1$, $e_2$, $e_3$, $e_4$
($e_r=|Z_r|, r=1,...,4$) and of the phase $\alpha$. The moduli $e_r$ of the skew-eigenvalues
are related to the following $\rm SU(8)$ invariants:
 \begin{eqnarray}
\left\{\matrix{
 I_1 &\equiv& \frac 12 Z_{AB} \bar Z^{AB} =(e_1)^2 + (e_2)^2+ (e_3)^2+
(e_4)^2\hfill \cr
 I_2 &\equiv& \frac 12 Z_{AB} \bar Z^{BC}Z_{CD} \bar Z^{DA}=(e_1)^4 +
(e_2)^4+ (e_3)^4+ (e_4)^4\hfill\cr
 I_3&\equiv&  -\frac 12 Z_{AB} \bar Z^{BC}Z_{CD} \bar
Z^{DE}Z_{EF} \bar Z^{FA}=(e_1)^6 + (e_2)^6+ (e_3)^6+ (e_4)^6\hfill\cr
 I_4 &\equiv & \frac 12
Z_{AB} \bar Z^{BC}Z_{CD} \bar Z^{DE}Z_{EF} \bar Z^{FG}Z_{GH} \bar Z^{HA}=(e_1)^8 + (e_2)^8+
(e_3)^8+ (e_4)^8\,.\hfill}\right.
\end{eqnarray}
The differential relations among the dressed charges still have a simple expression when
written in terms of the skew-eigenvalues. We find indeed:
\begin{eqnarray}
\left\{\matrix{\nabla_r e_1 &=& \Re\left[\e^{-\ii\frac \alpha 2} \left(P_{1,r}e_2 + P_{2,r} e_3+ P_{3,i} e_4\right)\right]\cr
\nabla_r e_2 &=& \Re\left[\e^{-\ii\frac \alpha 2} \left(P_{1,r}e_1 + P_{2,r} e_4+ P_{3,i} e_3\right)\right]\cr
\nabla_r e_3 &=& \Re\left[\e^{-\ii\frac \alpha 2} \left(P_{1,r}e_4 + P_{2,r} e_1+ P_{3,i} e_2\right)\right]\cr
\nabla_r e_4 &=& \Re\left[\e^{-\ii\frac \alpha 2} \left(P_{1,r}e_3 + P_{2,r} e_2+ P_{3,i} e_1\right)\right]}\right. \label{de8}
\end{eqnarray}
where $P_{1,i}=P_{1234,i}$, $P_{2,i}=P_{1256,i}$, $P_{3,i}=P_{3456,i}$ are the components of
the scalar vielbein appearing in \eq{de8} when  the central charge is written in normal form.

We then propose, for the prepotential $W$, the form
\begin{eqnarray}
W=a_1 e_1 + a_2 e_2 +a_3 e_3 + a_4 \e_4 \,,\label{W8}
\end{eqnarray}
giving, for its holomorphic gradient in normal form:
\begin{eqnarray}
\nabla W &=& \Re\Bigl\{P_{1}\left[\e^{-\ii\frac\alpha 2}( a_1 e_2 +a_2 e_1 )+ \e^{\ii\frac\alpha 2}(a_3 e_4 +a_4 e_3)\right]+\nonumber\\
&&+P_{2}\left[[\e^{-\ii\frac\alpha 2}(a_1 e_3 +a_3 e_1) +\e^{\ii\frac\alpha 2}(a_2e_4 +a_4 e_2)\right]+\nonumber\\
&&+P_{3}\left[[\e^{-\ii\frac\alpha 2}(a_1 e_4 +a_4 e_1) + \e^{\ii\frac\alpha 2}(a_2 e_3 +a_3
e_2)\right]\Bigr\}\,,\label{dW8}
\end{eqnarray}
where $P_{1},P_{2},P_{3}$ are the scalar vielbein 1-forms in normal form. Using \eq{diffrel8},
eq. \eq{W8} gives, for the right-hand side of \eq{potextr}:
\begin{eqnarray}
W^2+4\,g^{i\bar\jmath}\partial_iW\partial_{\bar\jmath}W&=& (a_1^2 + a_2^2 + +a_3^2 + a_4^2)\left(e_1^2 +e_2^2+e_3^2 +e_4^2\right) +\nonumber\\
&&+  2e_1e_2 (a_1a_2 + a_3 a_4 \cos\alpha )+ 2 e_3 e_4 (a_1a_2 \cos\alpha + a_3 a_4 )+ \nonumber\\
&&+e_1\to e_2 \to e_3 \to e_4 \to e_1\,.
\end{eqnarray}
This reproduces the result for the black-hole potential of supersymmetric theories,
\begin{equation}
V=I_1 = e_1^2 + e_2^2 + e_3^2 + e_4^2 \,,
\end{equation}
if
\begin{eqnarray}
\left\{\matrix{a_1^2 + a_2^2 + +a_3^2 + a_4^2&=&1\hfill\cr a_1a_2 + a_3 a_4 \cos\alpha
&=&0\hfill\cr a_1a_2  \cos\alpha+ a_3 a_4  &=&0\hfill\cr a_1a_3 + a_2 a_4 \cos\alpha
&=&0\hfill\cr a_1a_3\cos\alpha + a_2 a_4   &=&0\hfill\cr a_2a_3 + a_1 a_4 \cos\alpha
&=&0\hfill\cr a_2a_3\cos\alpha  + a_1 a_4  &=&0\,.\hfill}\right. \label{sys8}
\end{eqnarray}
This system is formally equivalent to (\ref{sys6}) if we replace $a_4$ with the singlet coefficient $b$ and
if the  overall  phase in $Z_{AB}$  is reinterpreted as the $N=6$ singlet phase. The solutions can be found
following the $N=6$ approach, paying attention to the fact that since here we consider the four charges on
the same footing, the BPS and the first non-BPS solutions of the $N=6$ case become equivalent in the $N=8$
version and correspond both to the BPS solutions of the $N=8$ case. Therefore now the system allows only for
two inequivalent solutions.
\begin{enumerate}
\item All the $a_i's$ vanish except one, say $a_1$: $a_1=1$, $a_2=a_3 =a_4=0$.

 In this case the prepotential encoding the solution has the
form:
\begin{eqnarray}
W_{(1)}&=& e_1\,,
\end{eqnarray}
which is extremized (see \eq{dW8}), for $e_1=e$, $e_2=e_3=e_4=0$. The Bekenstein--Hawking
entropy is:
\begin{eqnarray}
V_{(1)}|_{extr}= e^2\,.
\end{eqnarray}
This is the BPS solution.
\item $a_1=a_2=a_3=a_4= \frac 12$.

This solution requires that  the overall phase of the central charge in the normal form be
fixed to $\alpha=\pi$. We then have
\begin{eqnarray}
\left\{\matrix{W_{(2)}&=& \frac12 (e_1+e_2 + e_3+ e_4) \cr
\alpha&=&\pi \hfill}\right.
\end{eqnarray}
which is extremized  for
$e_1=e_2=e_3 =e_4=e$, $\alpha =\pi$.
 In this case the entropy is
\begin{eqnarray}
V_{(2)}|_{extr}=4 e^2
\end{eqnarray}
This is a non-BPS extremal solution.
\end{enumerate}


\subsubsection{The $N=2$ case}\label{n=2}

The vector multiplet moduli space of $N=2$ theory  is given by special geometry, and allows $\sigma$-models which are not in general homogeneous spaces. Since our ansatz for $W$ is given in terms of invariants of the representation of the isotropy group of the scalar manifold which the dressed charges belong to, a general result here is not easy to obtain. However, we know that the scalar manifold is embedded in the coset
$\cM_{SK} \subset Sp(2n+2)/U(n+1)$ due to the symplectic embedding. The $n+1$ dressed charges $(Z,Z_i)=\mathcal{Z}_A$ compose a vector of $U(n+1)$.
For a general special manifold there are many ($U(n+1)$) invariants (the only request is to build coordinate invariants of K\"ahler weight zero) which can be constructed in terms of $Z$ and $Z_i$ out of $g_{i\jbar}$, $C_{ijk}$, and their derivatives and products.

For special manifolds which are coset spaces $G/H$, the invariants are built in terms of  the invariant tensors of $H$, and the result is found by a case by case inspection.

In the minimal case $SU(1,n)/U(n)$, where the $C_{ijk}\equiv 0$, the procedure to find the prepotential is straightforward. Indeed in this case we only have the 2 possible invariants $e^2= Z\bar Z$ and $\rho^2= Z_i \bar Z_\jbar g^{i\jbar}$, so that
\begin{eqnarray}
W =a e + b \rho .\label{n=2c=0}
\end{eqnarray}
Using the differential relations of special geometry on \eq{n=2c=0} we find
\begin{eqnarray}
V_{BH}=(a^2 + b^2)(e^2 + \rho^2) + 2ab e\rho
\end{eqnarray}
which coincides with the supersymmetric one
\begin{eqnarray}
V_{BH}=|Z|^2 + Z_i \bar Z_\jbar g^{i\jbar}
\end{eqnarray}
for
\begin{eqnarray}
a^2 + b^2&=&1\nonumber\\
ab&=&0
\end{eqnarray}
This system has two independent solutions corresponding to the two attractors:
\begin{enumerate}
\item $a=1$, $b=0$.
In this case the prepotential encoding the solution has the form
$W_{(1)}= e
$
which is extremized  for $Z_i=0$.
The Bekenstein--Hawking entropy is
\begin{eqnarray}
V_{(1)}|_{extr}= e^2
\end{eqnarray}
This is the BPS solution.
\item $a=0$, $b=1$.
$W_{(2)}= \rho
$
which is extremized  for
$Z=0$.
 In this case the entropy is
\begin{eqnarray}
V_{(2)}|_{extr}= \rho^2
\end{eqnarray}
This is a non-BPS extremal solution.
\end{enumerate}

As a second example we may analyze the special manifold $SO^*(12)/U(6)$ from the $N=2$ point of view. In
this case the flat index $I$ is identified with the antisymmetric couple $AB$, and the $C$-tensor, with flat
indices is identified with the invariant tensor $\epsilon_{ABCDEF}$ \cite{urevisited}.   We expect the
following invariants to be involved: $Z\bar Z$, $Z_I\bar Z^I$, $\bar Z^J \bar Z^K Z_L Z_M C_{IJK}C^{I LM}$,
$|Z_IZ_J Z_K C^{IJK}|^2$.
They are in fact related to the invariants of $\rm SO^*(12)$ introduced in \eq{i6} by:
 \begin{eqnarray}
\left\{\matrix{ Z_I\bar Z^I &=& I_1\hfill\cr Z\bar Z &=& I_2\hfill
\cr \bar Z^J \bar Z^K Z_L Z_M C_{IJK}C^{I LM}&\propto& \left(I_1^2
-I_3\right)\hfill\cr |Z_IZ_J Z_K C^{IJK}|^2&\propto& \left(I_1^3
-3I_1 I_3 + 4 I_4\right)\,.\hfill}\right.\label{i2-6}
\end{eqnarray}
We may then give for the prepotential exactly the same Ansatz \eq{W6} as for the $N=6$ case with $e_r$ and $\rho$
 given in terms of the $N=2$ invariants \eq{i2-6} through \eq{i6}, finding exactly the same solutions
 as in section \ref{n=6}. As already discussed  there, the only difference is that, since in the
  $N=2$ interpretation the singlet $X$ is in fact the central charge while the $Z_{AB}=Z_I$ are the
    matter charges, the meaning of the first two attractor solutions enumerated in section \ref{n=6} is now interchanged: the BPS one is the second solution, while the first is now non-BPS.

Other examples are given in \cite{anna}.


\section{Concluding remarks and speculations on the non extremal case}\label{nonextremal}

In this paper we have dealt with the problem of finding, in four dimensional  extended supergravity, the analogue, for non-BPS extremal black holes, of the first order differential equations which encode the attractor mechanism for BPS black holes and which imply the second order field equations.

We have given a general Ansatz for the prepotential $W$ which
reproduces all the known attractors in $N\geq 3$ extended
supergravity.

\par In this concluding section, we discuss a
possible extension of our analysis to the non extremal case $c\neq
0$. In this more general situation, we may argue that a possible
generalization of the expression for
 the prepotential
$W$
might include an explicit dependence on the evolution parameter $\tau$, that is:
\begin{eqnarray}
\dot U = W(\Phi,\tau)\,e^U\,.
\label{udotnonextr}
\end{eqnarray}
Indeed, differentiating \eq{udotextr} with respect to $\tau$,
 we find:
\begin{equation}
\ddot U=(\dot U)^2 +\dot W e^U \,,\label{u:nonextr}
\end{equation}
where now:
\begin{equation}
\dot W = \dot \Phi^r \partial_r W  + \partial_\tau W \,.\label{Wdotnonextr}
\end{equation}
The (on-shell) expression for $V_{BH}$ is still formally the same
as for the extremal case:
\begin{equation}
V_{BH}=W^2 + e^{-U} \dot W.\label{pot1nonextr}
\end{equation}
However, when inserted in \eq{bhconstr} the above expression now gives (using \eq{geoeq'}):
\begin{equation}
\ddot U-(\dot U)^2 =\left( \dot \Phi^r \partial_r W  + \partial_\tau W\right)\, e^U = \frac 12 g_{rs} \dot \Phi^r \dot \Phi^s -c^2 \label{constr2nonextr} \,.
\end{equation}

For $\dot \Phi^r \neq 0$, eq. \eq{constr2nonextr} admits the particular solution
\begin{eqnarray}
\left\{\matrix{
\dot\Phi^r &=& 2\,e^U\, g^{rs}\,\partial_s W \cr
\partial_\tau W &=&-c^2 e^{-U} }\right. \,\label{nonextr}.
\end{eqnarray}
The integration of the second equation in \eq{nonextr} gives
\begin{eqnarray}
W^2(\Phi,U)=c^2 e^{-2U} +W^2_0(\Phi)\,.\label{neint}
\end{eqnarray}

Eq. \eq{nonextr} reproduces the correct description  of   non-extremal black holes near the
horizon. Indeed, given the general form \eq{dstau} for the space-time metric, for $\tau \to -\infty$ the first order correction to a generic charged
black-hole solution is
\begin{equation}
e^{-2U} \sim \frac A{4\pi} \left(\frac{\sinh (c\tau)}c\right)^2
\end{equation}
so that
\begin{eqnarray}
W = -\frac{d}{d\tau}e^{-U} = -\cosh(c\tau)\sqrt{\frac A{4\pi}}\label{wne}
\end{eqnarray}
giving
\begin{eqnarray}
\partial_\tau W =  -c\sinh(c\tau)\sqrt{\frac A{4\pi}} = -c^2 e^{-U}\,.
\end{eqnarray}
Note that \eq{wne} may also be written as
\begin{eqnarray}
W^2 =\frac A{4\pi}\left(1 + \sinh^2(c\tau)\right) = \frac A{4\pi} + c^2 e^{-2U}
\label{wne2}
\end{eqnarray}
which coincides with \eq{neint} for $W_0^2 =\frac A{4\pi}$.

For the non-extremal cases where eq.s \eq{nonextr} hold, the field equations for the scalar sector are in fact still first order as for the extremal case.
To show this, it is however necessary to make a slight modification to the effective potential. Indeed, using  \eq{nonextr}  the effective potential reads
\begin{eqnarray}
V_{BH}&=&W^2+2\,g^{rs}\,\partial_rW\,\partial_sW -c^2
e^{-2U}\,.\label{pot2nonextr}
\end{eqnarray}
By inserting eq.s \eq{nonextr} in the second order evolution
equation for the scalars, eq. \eq{geoeq}, we actually find an
inconsistency. However, since  the expression \eq{pot1nonextr} for
the black-hole potential is an on-shell relation, any expression
for $V_{BH}$ given by
\begin{eqnarray}
V_{BH}&=&W^2+2\,g^{rs}\,\partial_rW\,\partial_sW   -c^2 e^{-2U} +
\alpha\,e^{-U}(\partial_\tau W + c^2 e^{-U})\,\label{pot3nonextr}
\end{eqnarray}
is equivalent to \eq{pot2nonextr}. If we redo the calculation of the field equations for the scalars \eq{geoeq} with the parametric expression \eq{pot3nonextr}, we find that for $\alpha=2$ it is automatically solved when we use the Ansatz \eq{nonextr} for $\dot\Phi^r$.
For all the cases where eq.s \eq{nonextr} hold,  we then have the following expression for  the effective potential in terms of $W$:
\begin{eqnarray}
V_{BH}&=&W^2+2\,g^{rs}\,\partial_rW\,\partial_sW +  2\partial_\tau
W e^{-U} + c^2 e^{-2U} \,\label{potnonextr}
\end{eqnarray}
By inserting the explicit expression \eq{neint} in \eq{potnonextr}
we find
\begin{eqnarray}
V_{BH}&=&W_0^2+2\,g^{rs}\,\partial_rW_0\,\partial_sW_0 -\frac{c^2
e^{-2U}}{W^2_0 + c^2 e^{-2U}}
2\,g^{rs}\,\partial_rW_0\,\partial_sW_0\,.\label{potnonextr2}
\end{eqnarray}
Note that the extrema of the prepotential $W$ do not extremize
$V_{BH}$, corresponding to the fact that for non-extremal black
holes, the attractor mechanism is not expected to be at work nor
the horizon to be a fixed point for the scalar fields.

However, the expression \eq{potnonextr2} for the effective
potential has the feature of containing an explicit dependence on
the evolution parameter $\tau$. Such behavior could be acceptable
for purely bosonic theories such as fake supergravity
\cite{fake,weinberg}. For black holes in supersymmetric theories
this clashes with the request that the effective potential be
identified with the general expression \eq{geopot}, where it
depends on $\tau$ only through the scalars fields. We then have to
assume that, in the supersymmetric case, \eq{potnonextr2} is
rigorously valid only for the ``double non-extremal'' cases of
constant scalars. For more general solutions, at a finite distance
from the black-hole horizon we then expect eq. \eq{nonextr} to
receive corrections. For completely general non-extremal cases, we
do not expect to have a first-order description in terms of a
prepotential.

\bigskip

As a final remark, let us recall that in the BPS case the effective lagrangian \eq{effect} may be written in
terms of a sum of squares, as discussed in \cite{fgk,anna}. We want to give a similar treatment for general
extremal black holes and for all the cases where \eq{nonextr} hold and the effective potential takes the
form \eq{potnonextr}. To this aim,   we consider the following quantity:
\begin{eqnarray}
K&=&(\dot U -W e^U)^2 + \frac 12 g_{rs}(\dot \Phi^r -2 g^{r\ell}\partial_\ell W)(\dot \Phi^s -2 g^{sm}\partial_m W) \geq 0\label{k}
\end{eqnarray}
Using \eq{potnonextr} we find:
\begin{eqnarray}
\nonumber\\
K&=&\cL_{eff} -R \label{K}\,,
\end{eqnarray}
where
\begin{eqnarray}
R =-\left[\frac{d}{d\tau}(2e^UW) +c^2\right] =- \frac{d}{d\tau}(2e^UW +c^2 \tau) \,.
\end{eqnarray}
Eq. \eq{K} implies that the effective lagrangian $\cL_{eff}$ is bounded from below:
\begin{eqnarray}
\cL_{eff} \geq -\frac{d}{d\tau}(2e^UW + c^2\tau ) \label{minimum0}
\end{eqnarray}
The extremum value $\cL_{eff}=R$ is realized on-shell for
\begin{eqnarray}
\left\{\matrix{\dot U &=&W e^U\hfill\cr
 \dot \Phi^r &=&2 g^{r\ell}\partial_\ell W\hfill\cr
 \partial_\tau W &=& -c^2 e^{-U}\hfill}\right. \label{minimum}
\end{eqnarray}
Under our hypothesis, \eq{nonextr}, eq.s \eq{minimum} are always
verified and \eq{minimum0} then implies, on-shell, that
\begin{eqnarray}
\cL_{eff} = -\frac{d}{d\tau}(2e^UW+ c^2 \tau)
\end{eqnarray}
is a topological quantity characterizing the extremal solution
\begin{eqnarray}
S_{on-shell}=\int_{-\infty}^0 d\tau \cL_{eff} = -[2\dot U + c^2
\tau]^{\tau = 0}_{\tau  \to -\infty} = -2(M_{ADM}-c) + 2 c^2
\tau|_{-\infty}\,.\nonumber\\&&
\end{eqnarray}
The non extremal infinite contribution from $c^2$ may be understood as a ``vacuum energy" contribution to the action.

Our result  generalizes the argument for  ``non-extremal but BPS
solutions'' discussed in \cite{weinberg} to cases where the scalar
fields have a non trivial radial evolution. In \cite{weinberg}, it
is shown that the effective two dimensional model describing the
non-extremal but BPS black hole has a supersymmetric completion
where the first order equations play the role of Killing spinor
equations, even if there is an obstruction to  the four
dimensional uplift of this effective  supersymmetric model. It
would be interesting to  perform the same analysis for the class
of non-extremal black holes defined by the first order equations
\eq{nonextr}. This is left to a future investigation.

\section*{Acknowledgements}
We thank Anna Ceresole, Gianguido Dall'Agata, Sergio Ferrara and Sandip Trivedi for interesting discussions.

 Work supported in part by the European Community's Human Potential
Program under contract MRTN-CT-2004-005104 `Constituents, fundamental forces and symmetries of the
universe', in which the authors are associated to Torino University, and from MIUR PRIN project 2005.

\end{document}